\documentclass{andromedaone}       

\journal{BSM}
\vol{2017}
\jyear{Egypt}
\pages{Hurgada} 

\usepackage{epsf}
\usepackage{amsmath,amssymb}
\numberwithin{equation}{section}

\received{xx January 2018}
\published{xx March 2018}

\def\beq{\begin{equation}}
\def\eeq{\end{equation}}
\def\beqn{\begin{eqnarray}}
\def\eeqn{\end{eqnarray}}

\newcommand{\cc}[2]{c{#1\atopwithdelims[]#2}}

\newcommand{\nn}{\nonumber}

\def\vol#1#2#3{ {\bf {#1}} ({#2}) {#3}}
\def\NPB#1#2#3{{\it Nucl.\ Phys.}\/ {\bf B#1}, #3 (#2)}
\def\PLB#1#2#3{{\it Phys.\ Lett.}\/ {\bf B#1}, #3 (#2)}

\def\PRD#1#2#3{{\it Phys.\ Rev.}\/  {\bf D#1}, #3  (#2)}
\def\PRL#1#2#3{{\it Phys.\ Rev.\ Lett.}\/ {\bf #1}, #3 (#2)}

\def\MODA#1#2#3{ {\it Mod.\ Phys.\ Lett.}\/ {\bf A#1}, #3 (#2)}

\def\IJMP#1#2#3{ {\it Int.\ J.\ Mod.\ Phys.}\/ {\bf A#1}, #3 (#2)}

\def\EJP#1#2#3{ {\it Eur.\ Phys.\ Jour.}\/ {\bf C#1}, #3 (#2)}
\def\JHEP#1#2#3{ {\it JHEP}\/ {\bf #1}, #3 (#2)}

\def\etal{{\it et al\/}}

\def\AEF{A.E. Faraggi}

\begin{document}

\title{Spinor-Vector Duality and BSM Phenomenology}

\author{Alon E. Faraggi\auno{1}}
\address{$^1$Department of Mathematical Sciences, University of Liverpool,
  Liverpool L69 7ZL, United Kingdom}

\begin{abstract}
  Spinor--Vector Duality (SVD) has been observed in worldsheet
  constructions of heterotic--string compactifications. Recently,
  its realisation in the effective field theory limit of string
  vacua in six and five dimensions has been investigated. The SVD
  has been used to construct a string model that allows for an extra
  family universal $U(1)$, with the standard $E_6$ embedding of its
  charges, to remain unbroken down to low scales. Anomaly cancellation
  of the extra $U(1)$ charges mandates the existence of additional matter
  states at the extra $U(1)$ breaking scale, which affects precision
  measurements of Standard Model parameters. I discuss the construction
  of non-supersymmetric sting vacua and ``modular maps'' akin to the
  spacetime supersymmetry map. Such ``modular maps'' provide a glimpse
  into the enormous symmetry structure underlying the entire space of
  perturbative string vacua that is yet to be uncovered. 

 
\end{abstract}

\maketitle


\section{Introduction}\label{intro}
The Standard Model of particle physics successfully accounts for most of
the observable data to date. The Standard Model is not the end of the road.
In the first place it contains too many parameters. In the Standard Model
itself we may count the 45 gauge charges; the 9 fermion masses and
4 CKM parameters; the 3 gauge couplings; the Higgs VEV and coupling;
and the strong CPX parameter; for a total of 45+19= 64 parameters. If one
adds neutrino charges and masses to the melee, as indicated by experiments,
the counting grows further.
One often hears that physicists crave evidence for physics beyond
the Standard Model. If those came in the form of new forces and
new particles that would entail increasing the number of parameters
required to account for observations. An alternative approach 
is to reduce the number of free parameters by
refining the mathematical models. Grand Unified Theories (GUTs)
make a step in that direction by embedding the Standard Model
states in multiplets of the grand unification group. Most appealing
in this regard is the embedding in $SO(10)$ GUT in which the fermion
multiplets are embedded in three spinorial 16 representations of $SO(10)$.
Hence, the number of gauge charge parameters is reduced from
$3\times 3\times 6= 54$ to one, being the number of 16 representations
required to embed the Standard Model states, plus right--handed neutrinos.
Grand Unified Theories, however, cannot be the end of the road either.
There are still too many ad hoc parameters, in particular in the
flavour sector. The origin of the basic flavour structure, the duplication
of the family multiplets and the parameters that determine their masses
and mixing, can only be sought by embedding the Standard Model in a
theory of quantum gravity.

Recently, experimental evidence for physics beyond the Standard Model
has been in the news, generating substantial excitement
\cite{LHCbExp, gminus2}. 
It should be stated that the merit and value of these, and other
experiments, is not in providing evidence
for physics beyond the Standard Model. Their value and merit
is in reducing the error bars of the measurements of the
basic experimental observables. The marvel of the experiments is
in the design, construction and delivery of the specified
experimental targets in energy, luminosity, and other variables.
If an experiment is able to reduce the experimental uncertainty of
the basic observable parameters, then it is celebrated
as great success and triumph of human curiosity and ingenuity.
Whether or not it discovers physics beyond the Standard Model
is not a measure of its success. One can go further and
propose that experimentalists should not care at all about
physics beyond the Standard Model. All experimentalists
have to do is to improve the measurements of the Standard Model
parameters. If physics beyond the Standard Model exists it will
appear as an inconsistency in using the Standard Model parameters
to paramatrise the observable data. 

String theory provides a self consistent framework to explore the
embedding of the Standard Model in quantum gravity. Its consistency
conditions dictate the existence of the gauge, matter and scalar
sectors that are observed in nature. Furthermore, string theory
predicts the existence of a finite number of degrees of freedom
required in a perturbatively finite theory of quantum gravity. 
Thus, for example, in the perturbative heterotic--string the
rank of the gauge group cannot exceed 22. In some guise, the additional
degrees of freedom, beyond those observed in the Standard Model,
can be interpreted as extra spacetime dimensions. Since extra
dimensions, beyond the four spacetime dimensions detected via the
gauge and gravitational interactions, are not seen, they need to be
hidden from observations. This is achieved by making the extra
dimensions sufficiently small, so as to avoid detection. In other
guises, the extra degrees degrees of freedom required by consistency,
are represented in terms of free or interacting worldsheet fields
propagating on the string worldsheet. The process of constructing
consistent string solutions gives rise to a myriad of possibilities.
Whereas in ten dimensions the number of consistent theories
is relatively scarce, being five supersymmetric and eight
non--supersymmetric string theories, the number of consistent
solutions in four spacetime dimensions is large.
The meaning and interpretation of the myriad of solutions is an
open question. Should they be regarded as states in an Hilbert space
of quantum gravity with some probability measure? Or does there exist
a yet unknown mechanism that selects dynamically a single solution?
These are open questions that at present cannot be addressed.
Our understanding of quantum gravity is not sufficiently advanced.
String theories provide effective probes to explore some of the
properties of quantum gravity and construct phenomenological
models. But string theories do not provide an axiomatic framework,
a la general relativity or quantum mechanics, for a fundamental formulation
of quantum gravity. String theories do provide an arena in which
we can explore how the parameters of the Standard Model arise
in a perturbative theory of quantum gravity.
To advance this program requires progress on the basic understanding
of string theories and string compactifications, as well as on the
constructions of phenomenological models and their relation to the
Standard Model and its extensions. 

The construction of phenomenological string models proceeds by studying
compactifications in the effective field theory limit of string theories
as well as by studying exact string theory solutions. Ultimately,
the predictions extracted from string theory will be confronted with
the experimental data using an effective field theory parameterisation.
The string solutions in this context provide the boundary conditions.
Thus, whereas in the field theory context the parameters can be arbitrary,
it is only within the context of string theory that they are constrained.
Within the field theory approach there is nothing that constrains the
number of parameters that we can add to fit the experimental data. If the
fit does not work with one set of parameters, just add another one.
The straitjacket imposed by the quantum gravity constraints limits
this freedom, albeit at the expense of having a myriad of a priori
viable string vacua.

A characteristic feature of string theory is the existence of
various perturbative and non--perturbative duality symmetries that relate
different string solutions. Thus, the heterotic--string $E_8\times E_8$
and $SO(32)$ in ten dimensions are related via a $T$--duality transformation
in compactification to nine dimensions. Another celebrated example is
mirror symmetry that exchanges the complex and K\"ahler structure moduli
of the internally compactified complex manifold, and consequently reverse the
sign of the Euler characteristic. In this talk I will discuss another duality
that has been observed in heterotic--string compactifications under the
exchange of the total number of spinorial plus anti-spinorial
representations of the unbroken GUT group, and is dubbed spinor--vector
duality \cite{so10c10, fkrneq2, cfkr, aft, ffmt, fgh}.

The existence of the variety of duality symmetries in the space of
string compactifications is of paramount importance. From the worldsheet
point of view various duality symmetries can be realised in terms
of discrete torsions in the one--loop partition function.
It is seen that physical theories that are entirely distinct from
the point of view of the effective field theory limit are connected
in string theory. The reason is apparent. The string has access to its massive
modes, which are not accessible in the effective field theory limit,
and the duality transformations are induced by exchanging massless and
massive string modes. Furthermore, from the point of view of the string
theory, the realisation of dualities as exchange of discrete torsions
reflects modular properties of the one--loop partition function. From
the point of view of the low energy field theory description the duality
symmetries reflect an imprint of these modular properties in the
effective field theory representation of the string vacua.

One example where this picture is realised is in the case
of mirror symmetry on $Z_2\times Z_2$ orbifold, it was demonstrated
that the mirror map is induced by an exchange of a discrete torsion
\cite{vafawitten}. Mirror symmetry has profound implications 
on the internal complex manifolds that are utilised in the
effective field theory limit of string compactifications. It is
therefore anticipated that the rich modular properties of
the string worldsheet formalism may have similar profound implications
that are yet to be uncovered. Recently, we pursued this line of
inquiry in the case of spinor--vector duality, by exploring the
implications of the duality in the resolved limit of orbifold
compactifications \cite{fgh}.

While the duality symmetries inform us about the fundamental structure
of string theory, in particular, and quantum gravity, in general, they
may also have phenomenological implications. The non--vanishing of
neutrino masses is by now an established fact, Beyond the Standard Model.
Are they Dirac or Majorana? Are there additional light
states associated with the neutrino mass terms and that appear as
light sterile neutrinos? These are questions that will hopefully
be explored in forthcoming experiments. The question of the
existence of light sterile neutrinos is particularly interesting. There have
been some experimental indication for their existence, though
it would be fair to say that these do not look very convincing.
Albeit, possible existence of sterile neutrinos is an enigma from the
point of view of string phenomenology. What protects them from
acquiring mass terms of the order of the string or Planck scale?
One possible answer is that their lightness is protected by an additional
gauged $U(1)$ symmetry that remains unbroken down to relatively
low scale, and under which the sterile neutrinos are chiral \cite{steriles}.
Their
mass terms are then associated with this extra $U(1)$ breaking scale.
Just as the Standard Model chiral generation mass terms are associated
with the electroweak scale symmetry breaking.

Constructing heterotic--string models that allow for an extra $U(1)$ to
remain unbroken down to low scales turns out to be a non--trivial task. 
Heterotic--string constructions give rise to $SO(10)$ or $E_6$ embedding
of the Standard Model spectrum, and those do give rise to extra
$U(1)$ symmetries. In $SO(10)$ we have the gauged $U(1)_{B-L}$ and
$U(1)_{R}$ \cite{so10zprime}, whereas in $E_6$ we have an additional
family universal $U(1)$ symmetry \cite{u1a}. Additional, 
$U(1)$ symmetries that do not have a GUT embedding \cite{none6zprime}
can arise from the hidden sector or from the
compactified internal space. These may be family universal or
non--universal.
The focus in this talk is on extra $U(1)$ symmetries that have an $E_6$
embedding. To induce the seesaw mechanism
one of these $U(1)$ combinations has to be broken at an intermediate
or high energy scale. On the other hand, the symmetry breaking
of $E_6\rightarrow SO(10)\times U(1)_A$ in the string constructions
entails that the $U(1)_A$ symmetry is anomalous \cite{u1a}, and hence cannot be
part of a low scale $Z^\prime$.

The problem is therefore the construction of heterotic--string
models with anomaly free extra $U(1)\in E_6$. One route is to embed the
extra $U(1)$ in a non--Abelian symmetry, via the symmetry breaking
pattern $E_6\rightarrow SU(6)\times SU(2)$ \cite{e8patterns}.
This requires the breaking of the non-Abelian symmetry in the
effective field theory limit of the string model. 
The second route utilises the extraction of self--dual models under
spinor--vector duality, in which $U(1)_A$ is anomaly free
\cite{frzprime}. To understand
how this comes about, it is instrumental to examine the case of the
models with $E_6$ symmetry.
In this case $U(1)_A$ is anomaly free by virtue of its embedding in $E_6$.
The chiral representations in these models
are the $27$ and $\overline{27}$ of $E_6$, their decomposition under
$SO(10)\times U(1)_A$ is
\beqn
27 & = & 16_{\frac{1}{2}}+ 10_{-1}+1_{2} \nonumber\\
\overline{27} & = & \overline{16}_{-\frac{1}{2}}+ 10_{1}+1_{-2}.  \nonumber
\eeqn
Therefore, in the case of $E_6$ the total number of $16\oplus\overline{16}$
representations is equal to the total number of vectorial $10$ representations.
The $E_6$ models are self--dual under the spinor--vector duality.
This is similar to the case of $T$--duality on circle, where at the
self--dual point the symmetry is enhanced from $U(1)$ to $SU(2)$.
We can have string models with self--dual spectrum under spinor-vector
duality but without enhancement of the gauge symmetry to $E_6$. 
In such models the $U(1)_A$ can be anomaly free because the spectrum preserves
its $E_6$ embedding. This is possible in $Z_2\times Z_2$ orbifold when
the spinorial and vectorial components are obtained at different fixed
points, thus allowing the spectrum to preserve spinor--vector self--duality,
without enhancement of the gauge symmetry. 

\section{Fermionic $Z_2\times Z_2$ orbifolds}\label{fz2z2or}

Since the late eighties the heterotic--string models in the free fermionic
formulation \cite{fff} provided an arena to study the phenomenology of the
Standard Model and its Grand Unified extensions in a theory of
quantum gravity. Among those is the construction of the first
heterotic--string models that gave rise solely to the spectrum
of the Minimal Supersymmetric Standard Model in the effective
field theory limit \cite{fny}; the calculation of the
heavy generation Yukawa couplings and the prediction of
the top quark mass at $\sim175-180{\rm GeV}$ \cite{tqmp}
several years prior to its experimental observation \cite{topdiscovery};
mass and mixing matrices of the Standard Model quark and charged leptons 
\cite{fermionmasses}, as well as
left--handed neutrino masses via a generalised seesaw mechanism \cite{nmasses}; 
threshold corrections and string gauge coupling unification \cite{gcu};
proton lifetime \cite{ps}; 
supersymmetry breaking and squark degeneracy \cite{sd}; 
moduli fixing \cite{moduli}; and more \cite{more}. 

\begin{figure}[h]
\centering
\includegraphics[width=14pc]{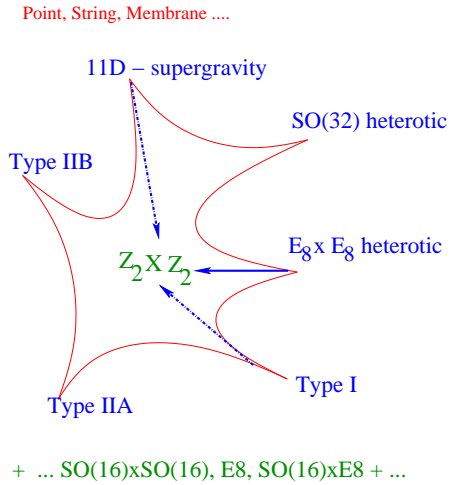}\hspace{2pc}%
\begin{minipage}[b]{14pc}
\caption{\label{mtheorypicture}%
\emph{
Perturbative treatment of elementary particles characterises 
them as idealised points, strings or membranes, ... 
The non--perturbative dualities 
of supersymmetric string theories 
in ten dimensions, as well as 11 dimensional supergravity
suggests probing the properties of different classes of string 
compactifications in the different limits. Uncovering the
string dynamics will necessitate bringing the non--supersymmetric
vacua into the fold as well. 
}
}
\end{minipage}
\end{figure}

The free fermionic models correspond to
$Z_2\times Z_2$ toroidal orbifolds at special 
points in the moduli space \cite{z2xz2}.
The untwisted moduli space of the symmetric orbifolds consist of
3 complex and 3 K\"ahler moduli, and is common in all the symmetric
$Z_2\times Z_2$ orbifolds. Assignment of asymmetric boundary
conditions allows for the projection of some or all of the
untwisted moduli. The twisted moduli vary between models. In models
with $(2,2)$ worldsheet supersymmetry the twisted moduli are matched
with the number of chiral and anti--chiral generations. In models
in which the $(2,2)$ worldsheet supersymmetry is broken to $(2,0)$
this association is no longer apparent and the twisted moduli fields are
mapped to charged fields in the massless string spectrum. This is
of vital importance for the phenomenology of the models and for
extracting the smooth effective field theory limit. Thus, it may be that
in some configurations the resolved limit cannot sustain an unbroken
Standard Model gauge group, because the fields needed for the singularity
resolutions are necessarily charged under the Standard Model group,
whereas in other vacua the twisted moduli may be mapped to fields
that are charged under the hidden sector gauge group. In these cases
the orbifold singularities can be resolved without affecting the observable
gauge symmetry. In this context the free fermionic constructions are
particularly instrumental, because they do not assume any a priori structure.
This brings to the fore many discrete torsions that are turned
off in the orbifold construction, because those typically start off from
the $E_8\times E_8$ or $SO(32)$ heterotic--strings
in ten dimensions and
compactify to four dimensions on an orbifold of a six dimensional
toroidal lattice. The internal moduli and the Wilson line moduli in this
case are treated distinctly and the discrete torsions between them are
turned off. On the other hand, in the free fermionic models the internal
and Wilson line moduli are mingled together and the discrete torsions
between them appear as GSO projection coefficients in the one--loop
partition function. It should be emphasised though that this does
not mean that the free fermionic models are distinct. Every fermionic
$Z_2\times Z_2$ model can be realised as an orbifold model with the
appropriate discrete torsions turned on and vice versa.
As such the free fermionic models are 
related to phenomenological studies of $Z_2\times Z_2$ orbifolds 
using other formalism, among those {\it e.g.} \cite{grootnibel}.
The $Z_2\times Z_2$ orbifold models represent a particular case and
other cases are studied \cite{others} as well,
using a variety of worldsheet and target space approaches.
The aim of string phenomenology is to develop
the tools to discern between the
different cases and identify their experimental signatures.
The perturbative and non--perturbative duality relations among ten dimensional
string vacua, as well as eleven dimensional supergravity \cite{ht}
shows, as illustrated in figure \ref{mtheorypicture},
that the different string theories are limits of a more fundamental
theory. This is an important lesson because it shows that theories
that look distinct from the point of view of the effective field theory
limit, are in fact related in string theory by various duality
transformations. Similarly, the observation of the spinor--vector
duality tells us that the myriad of string vacua with different
physical content in the effective field theory limit should not be
taken at face value. The dynamical picture in string theory may be very
different from what is indicated in the static limits. In this context,
it is also vital to explore not only the stable supersymmetric
configurations, but also non--stable configurations, 
compactified on the same underlying manifolds. As depicted in
figure \ref{mtheorypicture}, different limits should be compactified
on the same underlying manifold, being $Z_2\times Z_2$ orbifold in this
case study, as well as the non--supersymmetric and tachyonic
vacua, to explore the similarities and distinctions in the different
cases. It should be anticipated that non of the perturbative limits
can fully characterise the real vacuum. At best the perturbative
limits can provide effective probes that can capture some of its 
properties. For example, the embedding of the Standard Model states
in spinorial 16 representations of $SO(10)$ can only be gleaned in the
heterotic $E_8\times E_8$ string, because it is the only
limit that gives rise to spinorial representations in its
perturbative spectrum. On the other hand, the dilaton has a
run away behaviour in this limit and stabilising the dilaton necessitates
moving away from the perturbative $E_8\times E_8$ heterotic--string limit.

The $Z_2\times Z_2$ orbifold compactifications 
have been most extensively studied in the free fermionic formulation
of the heterotic-string in four dimensions.
This formulation is equivalent to the toroidal orbifold construction. 
For any free fermion model one can find the bosonic equivalent 
\cite{z2xz2}, and the two representations have their respective merits.
In particular, in the fermionic formulation many discrete torsions that
are a priori turned off in the orbifold constructions, appear as free
phases in the free fermionic models. This is particularly noted
in the case of the spinor--vector duality, which is induced by
discrete torsion between the orbifold $Z_2$ twist and the $Z_2$ Wilson
line that breaks $E_8\times E_8 \rightarrow SO(16)\times SO(16)$.
On the other hand, in the orbifold construction there is a clear
separation between the internal and Wilson line moduli, which is
blurred in the fermionic models. 
The free fermion formalism provides a robust framework to 
construct phenomenological string models and study their properties.

In the fermionic formulation of the heterotic--string in four dimensions
all the worldsheet degrees of freedom needed to cancel the conformal anomaly 
are represented in terms of two dimensional free fermions
on the string worldsheet. The 64 worldsheet 
fermions the lightcone gauge are denoted as:

\leftline{~~~${\underline{{\hbox{Left-Movers}}}}$:~~~~~~~~~~~~~~~~~~~~~~~~
~~~~$\psi^\mu,~~{ \chi_i},~~{ y_i,~~\omega_i}~~~~(\mu=1,2,~i=1,\cdots,6)$}
\vspace{4mm}
\leftline{~~~${\underline{{\hbox{Right-Movers}}}}$}
$${\bar\phi}_{A=1,\cdots,44}=
\begin{cases}
~~{ {\bar y}_i~,~ {\bar\omega}_i} & i=1,{\cdots},6\cr
  & \cr
~~{ {\bar\eta}_i} & i=1,2,3~~\cr
~~{ {\bar\psi}_{1,\cdots,5}} & \cr
~~{{\bar\phi}_{1,\cdots,8}}  & 
\end{cases}
$$
\noindent
Where $\{y,\omega\vert{\bar y},{\bar\omega}\}^{1,\cdots,6}$
correspond to the internal manifold six compactified dimensions; 
${\bar\psi}^{1,\cdots,5}$ produce the $SO(10)$ GUT symmetry; 
${\bar\phi}^{1,\cdots,8}$ generate the 
hidden sector gauge symmetry; and ${\bar\eta}^{1,2,3}$ 
produce three $U(1)$ symmetries in the observable sector.
Models in the free fermionic formulation are written in terms of a
set of boundary 
condition basis vectors, which denote the transformation properties 
of the fermions around the noncontractible loops of the vacuum to vacuum
amplitude, and the Generalised GSO (GGSO) projection coefficients of
the one loop partition function \cite{fff}.

\section{Classification of fermionic $Z_2\times Z_2$ orbifolds}
\label{classfz2z2or}

The early free fermionic models consisted of isolated examples
with a shared underlying GUT structure \cite{fsu5, fny, so64, lrs}. 
The basis vectors spanning the different cases contained the 
NAHE--set vectors \cite{nahe},
denoted as $\{ {\bf1}, S, b_1, b_2, b_3\}$. The NAHE--set gives rise
to an $SO(10)\times SO(6)^3\times E_8$ gauge symmetry, with forty--eight
multiplets in the spinorial {\bf 16} representation of $SO(10)$, arising
from the three twisted sectors of the $Z_2\times Z_2$ orbifold $b_1$, $b_2$
and $b_3$. The $S$--vector generates $N=4$ spacetime supersymmetry, 
which is reduced to $N=2$ by the basis vector $b_1$ and to $N=1$ by the 
inclusion of both $b_1$ and $b_2$.
The NAHE--set is augmented with three or four additional
basis vectors, typically denoted as $\{\alpha, \beta, \gamma\}$,
which break the $SO(10)$ gauge symmetry to one of its subgroups.
and simultaneously reduce the number of generations to three.
In the standard--like models \cite{fny} the 
$SO(10)$ gauge symmetry is broken 
to $SU(3)\times SU(2)\times U(1)_{B-L}\times U(1)_{R}$,
and the weak hypercharge is given by the combination 
$$U(1)_Y=T_{3_R} + {1\over2}(B-L)\in SO(10).$$
Each of the $b_1$, $b_2$ and $b_3$ sectors produces one generation
that form complete {\bf 16} multiplets of $SO(10)$. 
The models 
admit the needed scalar states to further reduce the gauge symmetry and
to produce a viable fermion mass and mixing spectrum 
\cite{tqmp,fermionmasses,nmasses}.

Since 2003, systematic classification of $Z_2\times Z_2$ heterotic--string
orbifolds have been developed using the free fermionic model building
rules. The classification method was initially developed for
the spinorial {\bf 16} and $\overline{\bf 16}$ representations in vacua
with unbroken $SO(10)$ gauge group \cite{so10class}, and subsequently
extended to include vectorial {\bf 10} representations \cite{so10c10}.
This led to the discovery of Spinor--Vector Duality (SVD) in the space of
fermionic $Z_2\times Z_2$ orbifold compactification, where the
duality transformation is induced by exchange of GGSO phases.
The classification method was subsequently extended
to vacua with:
$SO(6)\times SO(4)$ \cite{so64class}; 
$SU(5)\times U(1)$  \cite{fsu5class}; 
$SU(3)\times SU(2)\times U(1)^2$ \cite{slmclass}; 
$SU(3)\times U(1)\times SU(2)^2$ \cite{lrsranclass, lrsferclass},
unbroken subgroups of $SO(10)$.
In this classification method the string models are 
generated by a fixed set of boundary condition basis vectors, 
consisting of twelve to fourteen basis vectors, 
$
B=\{v_1,v_2,\dots,v_{14}\}.
$
The models with unbroken $SO(10)$ group are produced by a set of 
twelve basis vectors 
\begin{eqnarray}
v_1={\bf1}&=&\{\psi^\mu,\
\chi^{1,\dots,6},y^{1,\dots,6}, \omega^{1,\dots,6}~~~|
~~~\bar{y}^{1,\dots,6},\bar{\omega}^{1,\dots,6},
\bar{\eta}^{1,2,3},
\bar{\psi}^{1,\dots,5},\bar{\phi}^{1,\dots,8}\},\nonumber\\
v_2=S&=&\{\psi^\mu,\chi^{1,\dots,6}\},\nonumber\\
v_{3}=z_1&=&\{\bar{\phi}^{1,\dots,4}\},\nonumber\\
v_{4}=z_2&=&\{\bar{\phi}^{5,\dots,8}\},
\label{basis}\\
v_{4+i}=e_i&=&\{y^{i},\omega^{i}|\bar{y}^i,\bar{\omega}^i\}, \ i=1,\dots,6,
~~~~~~~~~~~~~~~~~~~~~N=4~~{\rm Vacua}
\nonumber\\
& & \nonumber\\
v_{11}=b_1&=&\{\chi^{34},\chi^{56},y^{34},y^{56}|\bar{y}^{34},
\bar{y}^{56},\bar{\eta}^1,\bar{\psi}^{1,\dots,5}\},
~~~~~~~~N=4\rightarrow N=2\nonumber\\
v_{12}=b_2&=&\{\chi^{12},\chi^{56},y^{12},y^{56}|\bar{y}^{12},
\bar{y}^{56},\bar{\eta}^2,\bar{\psi}^{1,\dots,5}\},
~~~~~~~~N=2\rightarrow N=1. \nonumber
\end{eqnarray}
The first ten basis vectors preserve $N=4$ spacetime supersymmetry.
The vectors $b_1$ and $b_2$ are $Z_2\times Z_2$ orbifold twists, and
the third twisted sector is obtained as the combination
$b_3= b_1+b_2+x$, where the $x$--sector is given by the combination
\beq
x= {\bf1} +S + \sum_{i=1}^6 e_i +\sum_{k=1}^2 z_k =
\{{\bar\psi}^{1,\cdots, 5}, {\bar\eta}^{1,2,3}\}.
\label{xmap}
\eeq
The breaking pattern $SO(10)\rightarrow SO(6)\times SO(4)$
is obtained by including in the basis the vector \cite{so64class}
\beq
v_{13}=\alpha = \{\bar{\psi}^{4,5},\bar{\phi}^{1,2}\},\label{so64bv}
\eeq
whereas $SO(10)\rightarrow SU(5)\times U(1)$ 
is achieved with the basis vector \cite{fsu5class}
\beq
v_{13}= \alpha = \{\overline{\psi}^{1,\dots,5}=\textstyle\frac{1}{2},
\overline{\eta}^{1,2,3}=\textstyle\frac{1}{2},
\overline{\phi}^{1,2} = \textstyle\frac{1}{2}, 
\overline{\phi}^{3,4} = \textstyle\frac{1}{2},
\overline{\phi}^{5}=1,\overline{\phi}^{6,7}=0,
\overline{\phi}^{8}=0\,\},\label{fsu5bv}
\eeq
and $SO(10)\rightarrow  SU(3)\times SU(2)\times U(1)^2$ is obtained
by adding (\ref{so64bv}) and (\ref{fsu5bv}) 
as two separate vectors, $v_{13}$ and $v_{14}$ to the basis \cite{slmclass}. 
The breaking of the $SO(10)$ gauge symmetry to the Left--Right Symmetric (LRS)
subgroup is obtained with the basis vector \cite{lrsranclass}, 
\begin{equation}
v_{13}=\alpha = \{ \overline{\psi}^{1,2,3} = \frac{1}{2} \; , \;
\overline{\eta}^{1,2,3} = \frac{1}{2}\; , \; \overline{\phi}^{1,\ldots,6} =
\frac{1}{2}\; , \; \overline{\phi}^7 \}.
\label{lrsbv}
\end{equation}
For a fixed set of basis vectors, 
the space of models is spanned by varying the independent
GGSO projection coefficients.
 For example, in the $SO(6)\times SO(4)$
models 66 phases are taken to be independent 
$$%
\bordermatrix{%
         &1   &  S  &  e_1  &   e_2   &  e_3   &  e_4  &   e_5  &   e_6%
&  z_1   &  z_2   &  b_1  &   b_2 & \alpha\cr%
   1   & -1   &  -1 &  \pm  &  \pm  &  \pm  &  \pm  &  \pm   &  \pm  & \pm  &%
 \pm   & \pm  & \pm & \pm\cr%
   S~   &   &    & -1 & -1 & -1 & -1 & -1 & -1 & -1 & -1 &  1 & 1 & -1\cr%
  e_1~  &   &    &    &\pm &\pm &\pm &\pm &\pm &\pm &\pm &\pm &\pm&\pm\cr%
  e_2~  &   &    &    &    &\pm &\pm &\pm &\pm &\pm &\pm &\pm &\pm&\pm\cr%
  e_3~  &   &    &    &    &    &\pm &\pm &\pm &\pm &\pm &\pm &\pm&\pm\cr%
  e_4~  &   &    &    &    &    &    &\pm &\pm &\pm &\pm &\pm &\pm&\pm\cr%
  e_5~  &   &    &    &    &    &    &    &\pm &\pm &\pm &\pm &\pm&\pm\cr%
  e_6~  &   &    &    &    &    &    &    &    &\pm &\pm &\pm &\pm&\pm\cr%
  z_1~  &   &    &    &    &    &    &    &    &    &\pm &\pm &\pm&\pm\cr%
  z_2~  &   &    &    &    &    &    &    &    &    &    &\pm &\pm&\pm\cr%
  b_1~  &   &    &    &    &    &    &    &    &    &    &    &\pm&\pm\cr%
  b_2~  &   &    &    &    &    &    &    &    &    &    &    &   &\pm\cr%
\alpha~ &   &    &    &    &    &    &    &    &    &    &    &   &   \cr%
  }, %
$$%
where the diagonal terms and below are fixed by modular invariance constraints.
The remaining fixed phases are set by requiring $N=1$ spacetime 
supersymmetry and the overall chirality.
Varying the 66 independent phases randomly scans a space
of $2^{66}$ (approximately $10^{19.9}$) $Z_2\times Z_2$ heterotic--string 
orbifold models.  A specific choice of the 66, $\pm1$ phases
corresponds to a distinct string vacuum with massless and massive
physical spectrum. The analysis proceeds by developing systematic tools
to analyse the entire massless spectrum, as well as the leading top quark
Yukawa coupling \cite{topyukawa}. Random choices of the 66 GGSO phases,
are generated and the spectrum is extracted systematically. A suitable
algorithm has to be implemented to ensure that identical models are
not generated. This is typically assisted by requiring that the
random cycle is sufficiently long and that the probability for the
generation of identical sequences is very small. It should be noted though
that the utility of the random generation method reaches its limit when
the space becomes to large \cite{slmclass,towardmac}. Genetic algorithm
provides a more efficient trawling tool to fish out models with specific
characteristics \cite{abelrizos}. On the other hand, genetic algorithms
are not suited to classifying and sorting large spaces of vacua. A more
suitable approach for that purpose is offered by application of
Satisfiability Modulo Theories, that can reduce the computer run time
by three orders of magnitude \cite{fpsw}. 

\section{Spinor--vector duality in heterotic--string orbifolds}\label{svdhsm}

\begin{figure}[h]
\includegraphics[width=14pc]{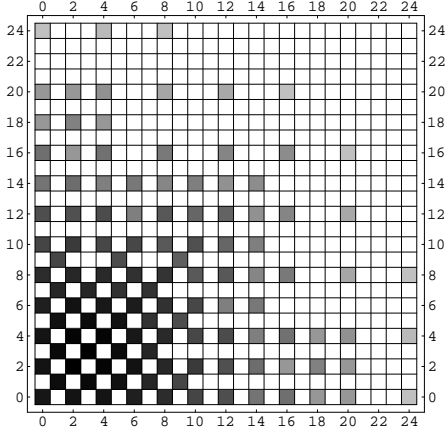}\hspace{2pc}%
\begin{minipage}[b]{14pc}
\caption{
\label{den}
Density plot showing the spinor--vector duality in the space of fermionic
$Z_2\times Z_2$ heterotic--string models. The plot shows the number 
of vacua with a given number of $({\bf 16}+\overline{\bf 16})$ and 
{\bf 10} multiplets of $SO(10)$. It is invariant under exchange of 
rows and columns, reflecting the spinor--vector duality 
underlying the entire space of vacua. Models on the diagonal are
self--dual under the exchange of rows and columns, {\it i.e.}
$\# ({\bf 16}+\overline{16}) = \#({\bf 10})$ without enhancement to
$E_6$, which are self--dual by virtue of the enhanced symmetry. 
}
\end{minipage}
\end{figure}

The free fermionic 
classification methodology led to the discovery of 
spinor--Vector Duality (SVD), depicted in figure \ref{den}, 
under the exchange of the total number of 
$({\bf 16}+\overline{\bf 16})$ spinorial and {\bf 10} 
vectorial representations 
of $SO(10)$ \cite{so10c10}.
The SVD arises from the breaking 
$(2,2)~ \rightarrow~ (2,0)$ worldsheet supersymmetry. It is
a general property of heterotic--string vacua.
From a worldsheet perspective, the SVD suggests that all string vacua
are connected by interpolations or by orbifolds, but are distinct
in the low energy effective field theory \cite{spwsp}.

\begin{figure}[!]
	\centering
	\includegraphics[width=100mm]{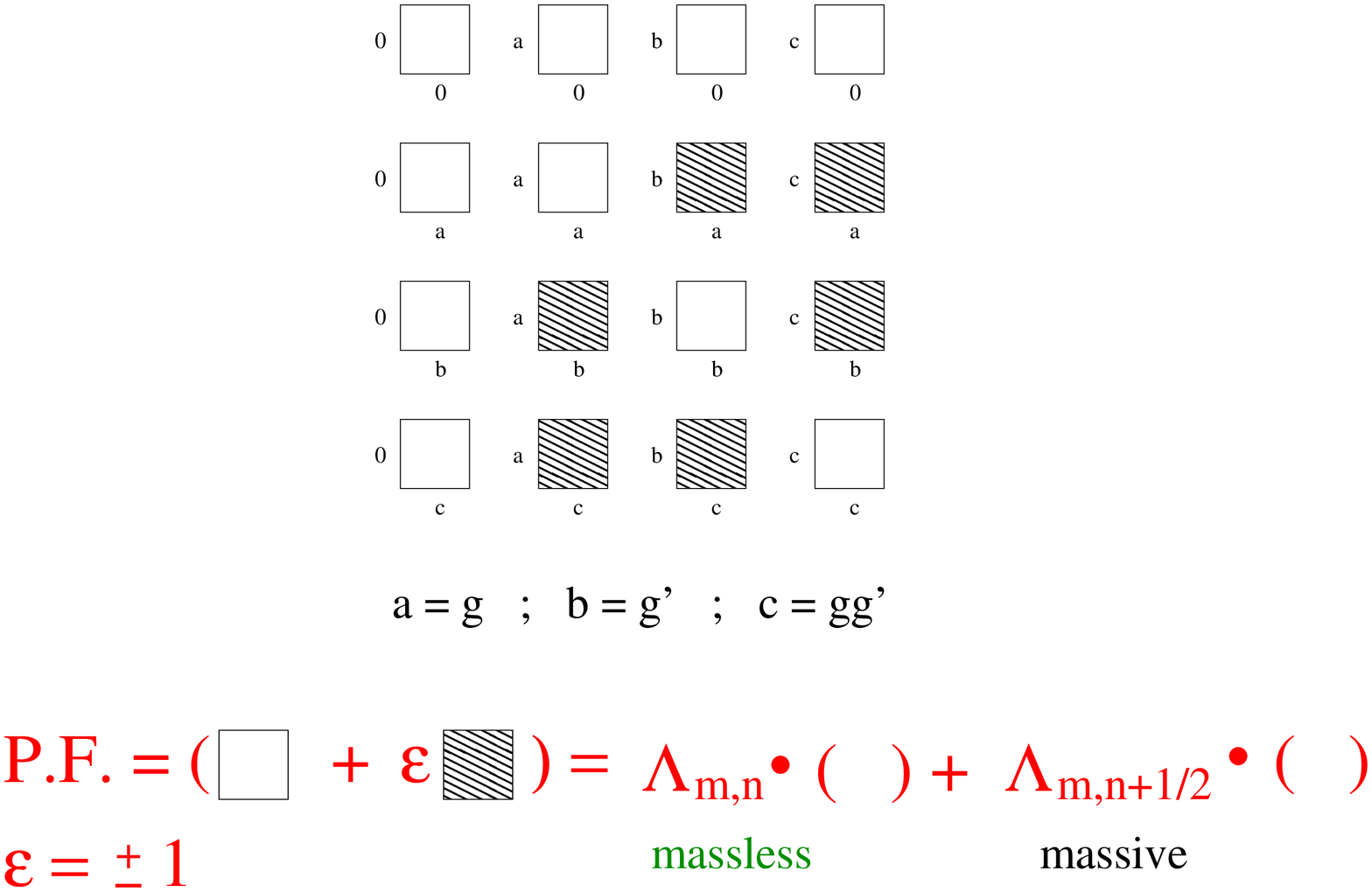}
	\caption{\emph{
            The
            $Z_2\times Z_2^\prime$ partition function of the $g$--twist and $g^\prime$ Wilson line
            with discrete torsion $\epsilon=\pm1$. 
}
}
	\label{z2z2svd}
\end{figure}

Further insight into the spinor-vector duality is obtained
by translating to the bosonic $Z_2\times Z_2$ representation.
First, it is noted that in the fermionic constructions the
SVD operates separately in each of the $Z_2$ planes, which
preserve $N=2$ spacetime supersymmetry. Hence, we can
study the SVD in vacua with a single $Z_2$ twist of the
compactified coordinates \cite{fkrneq2}.
Using the level one $SO(2n)$ characters
\beq
O_{2n} = {1\over 2} \left( {\theta_3^n \over \eta^n} +
{\theta_4^n \over \eta^n}\right) \,,
~~~~~~~~~~~~
V_{2n} = {1\over 2} \left( {\theta_3^n \over \eta^n} -
{\theta_4^n \over \eta^n}\right) \,,
~~~~~~~~~~~~
S_{2n} = {1\over 2} \left( {\theta_2^n \over \eta^n} +
i^{-n} {\theta_1^n \over \eta^n} \right) \,,
~~~~~~~
C_{2n} = {1\over 2} \left( {\theta_2^n \over \eta^n} -
i^{-n} {\theta_1^n \over \eta^n} \right) \,,
\label{so2nchara}
\eeq
{where} 
\beq
{
\theta_3\equiv Z_f{0\choose0}~~~
  \theta_4\equiv Z_f{0\choose1}~~~}
{
  \theta_2\equiv Z_f{1\choose0}~~~
  \theta_1\equiv Z_f{1\choose1}~,~~}\nonumber
\eeq
and $Z_f$ is the partition function of a single
worldsheet complex fermion, given in terms of 
theta functions \cite{manno},
the partition function
of the heterotic $E_8\times E_8$ string compactified to four
dimensions
\beq
{Z}_+ = (V_8 - S_8) \, 
\left( \sum_{m,n} \Lambda_{m,n}\right)^{\otimes 6}\, 
\left(\overline{ O} _{16} + \overline{ S}_{16} \right) 
\left(\overline{O} _{16} + \overline{ S}_{16} \right)\,,
\label{zplus}
\eeq
where as usual, for each circle,
\beq
p_{\rm L,R}^i = {m_i \over R_i} \pm {n_i R_i \over \alpha '} \,
~~~~~~~~~
{\rm and}
~~~~~~~~~
\Lambda_{m,n} = {q^{{\alpha ' \over 4} 
p_{\rm L}^2} \, \bar q ^{{\alpha ' \over 4} 
p_{\rm R}^2} \over |\eta|^2}\,.
\nonumber
\eeq
A $Z_2\times Z_2^\prime:g\times g^\prime$ action is applied.
The first $Z_2$ is freely acting. It
couples a fermion number in the observable and hidden sectors with 
a $Z_2$--shift in a compactified coordinate, and is given by
$
g: (-1)^{(F_{1}+F_2)}\delta
$
~where the fermion numbers $F_{1,2}$ act on the spinorial
representations of the observable and hidden $SO(16)$ groups as
$
F_{1,2}:({\overline O}_{16}^{1,2},
             {\overline V}_{16}^{1,2},
             {\overline S}_{16}^{1,2},
             {\overline C}_{16}^{1,2})\longrightarrow~
            ({\overline O}_{16}^{1,2},
             {\overline V}_{16}^{1,2},
             -{\overline S}_{16}^{1,2},
             -{\overline C}_{16}^{1,2})
$
~and $\delta$ identifies points shifted by a $Z_2$ shift 
in the $X_9$ direction, {\it i.e.} 
$
\delta X_9 = X_9 +\pi R_9.~
$
The effect of the shift is to insert a factor of $(-1)^m$ into the lattice 
sum in eq. (\ref{zplus}), {\it i.e.} 
$
\delta:\Lambda_{m,n}^9\longrightarrow(-1)^m\Lambda_{m,n}^9.
$
~The second $Z_2$ acts as a twist on the internal coordinates 
given by 
$
{g^\prime}:(x_{4},x_{5},x_{6},x_7,x_8,x_9)
\longrightarrow
(-x_{4},-x_{5},-x_{6},-x_7,+x_8,+x_9). 
$
Alternatively, the first $Z_2$ action can be interpreted as a Wilson line
in $X_9$ \cite{ffmt},
$
g~: (0^7,1|1, 0^7)  ~\rightarrow~
E_8\times E_8\rightarrow SO(16)\times SO(16).\nonumber
$
The effect of the single space twisting is to break $N=4\rightarrow N=2$ spacetime
supersymmetry and $E_8\rightarrow E_7\times SU(2)$
or with the inclusion of the Wilson line $SO(16)\rightarrow SO(12)\times SO(4)$.
The orbifold partition function is given by
$${Z~=~
\left({Z_+\over{Z_g\times Z_{g^{\prime}}}}\right)~=~
\left[{{(1+g)}\over2}{{(1+g^\prime)}\over2}\right]~Z_+}.$$
The partition function contains an untwisted sector and three twisted sectors.
It has the schematic form shown in figure \ref{z2z2svd}.
The winding modes in the sectors twisted by 
$g$ and $gg^\prime$ are shifted by $1/2$, and therefore these sectors only
produce massive states. The sector twisted by $g$
gives rise to the massless twisted matter states. 
The partition function has two modular orbits
and one discrete torsion $\epsilon=\pm1$. 
Massless states are obtained for vanishing lattice modes. 
The terms in the sector $g$ contributing to the massless 
spectrum take the form
\beqn
& &     \Lambda_{p,q}
\left\{
 {1\over2}
\left( 
           \left\vert{{2\eta}\over\theta_4}\right\vert^4
         +
           \left\vert{{2\eta}\over\theta_3}\right\vert^4
\right)
\left[{
       P_\epsilon^+Q_s{\overline V}_{12}{\overline C}_4{\overline O}_{16}} +
   {P_\epsilon^-Q_s{\overline S}_{12}{\overline O}_4{\overline O}_{16} }
\right.
{\left.  \right] + }
\right. \nonumber\\ 
& &\nonumber\\ 
& &\left.
~~~~~~~~~~{1\over2}\left(    \left\vert{{2\eta}\over\theta_4}\right\vert^4
                      -
                         \left\vert{{2\eta}\over\theta_3}\right\vert^4\right)
\left[{
P_\epsilon^+Q_s
{\overline O}_{12}{\overline S}_4{\overline O}_{16}} \right.
{\left. \right] } 
\right\}~~~~~~~~~~~~~~~~~~~~~~~~~~~~
+~~\hbox{massive}
 \label{masslessterminpf}
\eeqn
where 
\beq
P_\epsilon^+~=~\left({{1+\epsilon(-1)^m}\over2}\right)\Lambda_{m,n}~~~;~~~
P_\epsilon^-=\left({{1-\epsilon(-1)^m}\over2}\right)\Lambda_{m,n} 
\label{pepluspeminus}
\eeq
Depending on the sign of $\epsilon=\pm$ 
it is seen from  eq. (\ref{pepluspeminus}) that either the vectorial states, 
or the spinorial states, are massless. In the case with $\epsilon=+1$ 
we note from eq. (\ref{eplus}) that in this case massless momentum
modes from the shifted lattice arise in $P_\epsilon^+$ whereas 
$P_\epsilon^-$ only produces massive modes. Therefore, in his case  
the vectorial character ${\overline V}_{12}$ in eq. (\ref{pepluspeminus})
produces massless states, whereas the spinorial character
${\overline S}_{12}$ generates massive states.
In the case with $\epsilon=-1$ eq. (\ref{eminus}) shows
that exactly the opposite occurs. 
\beqn
{\epsilon~=~+1~~}&{\Rightarrow}&
{~~P^+_\epsilon~=~~~~~~~~~~~\Lambda_{2m,n}~~~
~~~~~~~~~~~~{ P^-_\epsilon~=~~~~~~~~
\Lambda_{2m+1,n}}~~~}\label{eplus}\\
{\epsilon~=~-1~~}&{\Rightarrow}&
{{ ~~P^+_\epsilon~=~~~~~~~~~~~\Lambda_{2m+1,n}}~~~
~~~~~~~~~P^-_\epsilon~=~~~~~~~~\Lambda_{2m,n}~~~}\label{eminus}
\eeqn
Thus, the spinor--vector duality is generated by the exchange of the discrete torsion
$\epsilon=+1\rightarrow \epsilon =-1$ in the $Z_2\times Z_2^\prime$ partition function.
This is very similar the the case of mirror symmetry in the $Z_2\times Z_2$
orbifold model of ref. \cite{vafawitten}, where the mirror symmetry map is induced
by exchange of the discrete torsion between the two orbifold $Z_2$ twists.
In the mirror symmetry case the chirality of the fermion multiplets is changed,
together with the exchange of the complex and K\"ahler moduli of the internal
manifold. The total number of degrees of freedom is invariant under the mirror
symmetry map. It is interesting to note that this is also the case in the
case of the SVD. In this case there is a mismatch between the number of states
in the vectorial, $12\cdot 2=24$, and spinorial $32$, cases. It is noted
from the 
second line in eq. (\ref{masslessterminpf}) that
the vectorial case $\epsilon=+1$
is accompanied by $8$ additional states, which are singlets
of the $SO(12)$ GUT group. It is seen that the total number of
degrees of freedom is preserved
under the duality map, {\it i.e.} 
$
{12\cdot 2+ 4\cdot2}{=}{32}.
$

What can be inferred from the spinor--vector duality? What lessons can we draw from the
example of mirror symmetry, which was initially observed in worldsheet string constructions?
Mirror symmetry has profound implications for the geometry of the internal manifold in the
effective field theory limit of the string compactifications. The SVD tells us that vacua
that look distinct from the point view of the effective field theory limit are
in fact connected in the worldsheet description, because in the string construction
massless and massive mode can be exchanged. Furthermore, the matching of the
massless degrees of freedom in the different cases tells us that the worldsheet
string theory primarily cares about obtaining the number of degrees of freedom
required in a modular invariant partition function, whereas how they are organised
in representations of the four dimensional gauge symmetry is of secondary importance. 
It is of further interest to explore the implications of the SVD in the effective
field theory limit of the string compactifications. The SVD can serve as a
probe of the moduli spaces of heterotic--string compactifications with 
$(2,0)$ worldsheet supersymmetry. While the moduli spaces of string
compactifications
with standard embedding and $(2,2)$ worldsheet supersymmetry
are fairly well understood,
the case of $(2,0)$ models is obscured. The SVD can provide
a very useful probe of
these models that in the string
picture can be seen as deformations of the $(2,2)$ cases, whereas in the
effective QFT picture they correspond to compactifications on Calabi-Yau
manifolds with vector bundles. Recently, we demonstrated the viability
of these approach in the case of compactifications to six and five dimensions
\cite{fgh}, where the effective field theory limit is obtained by resolving the
orbifold singularities. In this context, the worldsheet description serves as a
guide to guess how the discrete torsion of the worldsheet description should
be interpreted in the effective field theory limit. In this respect, it is
noted that the SVD in the worldsheet formalism generalises to string
compactifications
with interacting internal CFTs \cite{afg},
as well as to cases that include more discrete
torsions \cite{aft}.

\section{Low scale $Z^\prime$ in free fermionic models}\label{lsszprimeffm}

The interest in extra $Z^\prime$ vector bosons at low scales in string
derived models
stems from the role that the $U(1)$ symmetry can play in explaining
some of the
features of the Standard Model or the supersymmetric Standard Model.
This include
suppression of proton decay mediating operators and of the $\mu$--term
in supersymmetric models. However, construction of string models that allow
an extra $U(1)$ symmetry to remain unbroken down to intermediate or low scales
has proven to be non--trivial.
The first case to be considered
\cite{so10zprime}
was the combination 
\beq
U(1)_{Z^\prime}={3\over2}U(1)_{B-L}-2 U(1)_R\in SO(10), 
\label{u1zpinso10}
\eeq
which ensures suppression of proton decay from dimension four operators.
However, the underlying $SO(10)$ symmetry 
in the string models implies that the Dirac mass terms of the
tau neutrino and top quark are equal, necessitating the breaking 
of (\ref{u1zpinso10}) at high scale \cite{tauneutrino}.
A more natural possibility is that this 
$U(1)_{Z^\prime}$ symmetry is broken at a high scale, 
which generates a large scale seesaw and naturally produces 
light neutrino masses \cite{nmasses}.
Existence of alternative $U(1)$ symmetries 
in the string models that suppress the proton 
decay mediating operators
and allow a high seesaw mass scale were discussed in \cite{none6zprime}.
However, those tend to fail as low scale candidates, because they are
either non--family universal or have to be broken by the
supersymmetric $F$-- and $D$--flatness constraints.
Furthermore, the additional family universal $U(1) \in E_6$
is generically anomalous in the string models because of the symmetry
breaking pattern $E_6\rightarrow SO(10)\times U(1)$ that projects out
some of components of the chiral $27$ and $\overline{27}$
representations of $E_6$ \cite{u1a}. On the other hand,
gauge coupling unification
favours extra low scale $U(1)$ with an $E_6$ embedding of
its matter states \cite{viraf}.

The construction of string models that allow an extra $U(1)\in E_6$
to remain unbroken down to low scale can follow one of two routes.
The first is to use a different symmetry breaking pattern than the
$E_6\rightarrow SO(10)\times U(1)$ route. This symmetry breaking pattern
follows from the underlying breaking of $E_8\times E_8\rightarrow
SO(16)\times SO(16)$, which is commonly used in the free fermion models.
A different route would essentially correspond to keeping in the spectrum
the spacetime vector bosons from the $x$--sector, which enhances
$SO(16)\rightarrow E_8$. An example of a string derived model in this
class is the $SU(6)\times SU(2)$ model of ref. \cite{e8patterns}.

An alternative route, pursued in ref. \cite{frzprime}, is to use
the spinor--vector duality. As discussed above, in string models with
$E_6$ symmetry, the $U(1)_A\in SO(10)\times U(1)$ combination
is family universal by virtue of its embedding in $E_6$.
The $E_6$ representations are self--dual under the spinor--vector
duality and $U(1)_A$ is anomaly free. However, we can obtain
models which are self--dual under SVD without enhancement of the
$SO(10)\times U(1)$ gauge symmetry to $E_6$. In this case the
twisted matter representations still form complete $E_6$ multiplets,
which results in the universal $U(1)\in E_6$ combination being
anomaly free, without, however, enhancement of the gauge symmetry
to $E_6$.
A set of phases producing a model with the required properties 
is shown in eq. (\ref{BigMatrix}). 
The observable gauge group in the model is
$SO(6)\times SO(4)\times U(1)_{1,2,3}$ and the family universal 
combination,
$U(1)_\zeta=U(1)_1+U(1)_2+U(1)_3$,
is anomaly free.

\beq \label{BigMatrix}  (v_i|v_j)\ \ =\ \ \bordermatrix{
      & 1& S&e_1&e_2&e_3&e_4&e_5&e_6&b_1&b_2&z_1&z_2&\alpha\cr
 1    & 1& 1&  1&  1&  1&  1&  1&  1&  1&  1&  1&  1&      1\cr
S     & 1& 1&  1&  1&  1&  1&  1&  1&  1&  1&  1&  1&      1\cr
e_1   & 1& 1&  0&  0&  0&  0&  0&  0&  0&  0&  0&  0&      1\cr
e_2   & 1& 1&  0&  0&  0&  0&  0&  1&  0&  0&  0&  1&      0\cr
e_3   & 1& 1&  0&  0&  0&  1&  0&  0&  0&  0&  0&  1&      1\cr
e_4   & 1& 1&  0&  0&  1&  0&  0&  0&  0&  0&  1&  0&      0\cr
e_5   & 1& 1&  0&  0&  0&  0&  0&  1&  0&  0&  0&  1&      1\cr
e_6   & 1& 1&  0&  1&  0&  0&  1&  0&  0&  0&  1&  0&      0\cr
b_1   & 1& 0&  0&  0&  0&  0&  0&  0&  1&  1&  0&  0&      0\cr
b_2   & 1& 0&  0&  0&  0&  0&  0&  0&  1&  1&  0&  0&      1\cr
z_1   & 1& 1&  0&  0&  0&  1&  0&  1&  0&  0&  1&  1&      0\cr
z_2   & 1& 1&  0&  1&  1&  0&  1&  0&  0&  0&  1&  1&      0\cr
\alpha& 1& 1&  1&  0&  1&  0&  1&  0&  1&  0&  1&  0&      1\cr
  }
\eeq

The complete massless spectrum, as well as the tri--level 
superpotential are given in ref. \cite{frzprime}. 
The massless chiral spectrum in the model is self--dual under
the spinor--vector duality.
The model contains three chiral generations, as well as the required
heavy and light Higgs states to produce a realistic fermion
mass spectrum, as well as a cubic level top quark Yukawa coupling.
A VEV of the heavy Higgs field that breaks the Pati--Salam 
symmetry to the Standard Model along flat directions leaves the 
unbroken combination 
\beq
U(1)_{Z^\prime}~~=~~{1\over5}~U(1)_C~~-~~{1\over5}~U(1)_L-U_\zeta.   
\label{u1zpsdm}
\eeq
This $U(1)$ symmetry is anomaly free in this model and may
remain unbroken down to low scales.

\section{String inspired BSM phenomenology}\label{sizpp}

The string derived model gives rise to an extra $U(1)_{Z^\prime}$ combination
given in eq. (\ref{u1zpsdm}). There are numerous reasons that motivate the
possibility that this $U(1)$ symmetry remains unbroken down to low scales.
To explore the phenomenological implications of the model we can
choose the low scale spectrum to be consistent with the existence of the
$U(1)$ symmetry at low scales and impose some additional conditions
inspired from the string derived model. For example, we can fix the top
quark Yukawa coupling to be given by $\lambda_t= \sqrt{2}g$, where $g$ is the
gauge coupling at the heterotic--string unification scale \cite{tqmp}.
Similarly, the Yukawa couplings of the lighter quarks and leptons can
be calculated in the model from higher order non--renormalisable operators
\cite{fermionmasses} and detailed mass textures can be obtained.
Naturally, more details are subject to increasing model dependence and
requires making further assumptions, {\it e.g.} assumptions on a SUSY
breaking mechanism and SUSY breaking parameters. At this stage the
analysis is inspired from the string derived model and uses some input
parameters fixed by the string model. In this spirit, we can fix the
spectrum of the string inspired model, as given in table \ref{table27rot}.

\begin{table}[!ht]
  \tbl{
\it
Spectrum and
$SU(3)_C\times SU(2)_L\times U(1)_{Y}\times U(1)_{{Z}^\prime}$ 
quantum numbers, with $i=1,2,3$ for the three light 
generations. The charges are displayed in the 
normalisation used in free fermionic 
heterotic--string models. \label{table27rot}}
{\small
\begin{tabular}{|l|cc|c|c|c|}
\hline
Field &$\hphantom{\times}SU(3)_C$&$\times SU(2)_L $
&${U(1)}_{Y}$&${U(1)}_{Z^\prime}$  \\
\hline
$Q_L^i$&    $3$       &  $2$ &  $+\frac{1}{6}$   & $-\frac{2}{5}$   ~~  \\
$u_L^i$&    ${\bar3}$ &  $1$ &  $-\frac{2}{3}$   & $-\frac{2}{5}$   ~~  \\
$d_L^i$&    ${\bar3}$ &  $1$ &  $+\frac{1}{3}$   & $-\frac{4}{5}$  ~~  \\
$e_L^i$&    $1$       &  $1$ &  $+1          $   & $-\frac{2}{5}$  ~~  \\
$L_L^i$&    $1$       &  $2$ &  $-\frac{1}{2}$   & $-\frac{4}{5}$  ~~  \\
\hline
$D^i$       & $3$     & $1$ & $-\frac{1}{3}$     & $+\frac{4}{5}$  ~~    \\
${\bar D}^i$& ${\bar3}$ & $1$ &  $+\frac{1}{3}$  &   $+\frac{6}{5}$  ~~    \\
$H^i$       & $1$       & $2$ &  $-\frac{1}{2}$   &  $+\frac{6}{5}$ ~~    \\
${\bar H}^i$& $1$       & $2$ &  $+\frac{1}{2}$   &   $+\frac{4}{5}$   ~~  \\
\hline
$S^i$       & $1$       & $1$ &  ~~$0$  &  $-2$       ~~   \\
\hline
$h$         & $1$       & $2$ &  $-\frac{1}{2}$  &  $-\frac{4}{5}$  ~~    \\
${\bar h}$  & $1$       & $2$ &  $+\frac{1}{2}$  &  $+\frac{4}{5}$  ~~    \\
\hline
$\phi$       & $1$       & $1$ &  ~~$0$         & $-1$     ~~   \\
$\bar\phi$       & $1$       & $1$ &  ~~$0$     & $+1$     ~~   \\
\hline
\hline
$\zeta^i$       & $1$       & $1$ &  ~~$0$  &  ~~$0$       ~~   \\
\hline
\end{tabular}
}
\end{table}

It is noted that anomaly cancellation of the extra
$U(1)_{Z^\prime}$ symmetry requires the existence
of additional matter states at the $Z^\prime$ breaking scale.
These additional matter states may have profound implications on
experimental searches beyond the Standard Model, and may in fact
be associated with the recent observed deviations from the expected
Standard Model values in lepton universality \cite{LHCbExp} and
muon $g-2$ \cite{gminus2} experiments. A quick glance in table
(\ref{table27rot}) makes this evident. The model predicts the existence
of additional $SU(2)$ doublets and $SU(3)$ triplets that are chiral under
the extra $U(1)_{Z^\prime}$ but are vector--like with respect to the
Standard Model gauge group. The natural mass scale for these states
is the $U(1)_{Z^\prime}$ breaking scale. On the other hand, these additional
electroweak doublets and colour triplets are precisely the type of states
that may explain the deviations from the Standard Model predictions, via their
contributions in multi--loop diagrams.

Additionally, the existence of the extra $SO(10)$ singlet fields $S_i$ in
table \ref{table27rot} may explain the generation of the electroweak
scale, via dimensional transmutation. As seen in table \ref{table27rot},
gauge coupling unification at the heterotic--string scale suggests the
existence of an additional pair of Higgs doublets, beyond those that are
required by anomaly cancellation. This pair of additional doublets is
somewhat ad hoc, but beyond that, the traditional mixing term
between the chiral doublets is generated by the VEV of the $S_i$
fields. Thus, the low scale breaking of $U(1)_{Z^\prime}$ and the ensuing
electroweak symmetry breaking can be nicely incorporated in the $Z^\prime$ model.

From table \ref{table27rot} it seen that the model predicts the existence of
sterile neutrinos. Three sterile neutrinos are obtained from the Standard Model
singlets in the spinorial {\bf 16} representation of $SO(10)$. Mass terms for these
states are generated at the seesaw high mass scale, and they decouple from the
effective low scale field theory at that scale. Additionally, the model contain
the three $SO(10)$ singlet states $S_i$, which can appear as low scale
sterile neutrinos \cite{nmasses}. Existence of sterile neutrinos in this
model is correlated with the existence of a low scale extra $U(1)_{Z^\prime}$
symmetry that protects the sterile neutrinos from acquiring high scale mass.

\section{Spinor--vector duality and modular maps}\label{svdmm}

Modular maps are ubiquitous in string theory. By modular maps here I mean
maps that are induced by basis vectors with four periodic fermions in the
left-- or right--moving sector. An example of such a map is given by the
$S$--vector in eq. (\ref{basis}). The $S$--vector is the supersymmetry
generator in the string models and maps bosonic to fermionic sectors.
The spinor--vector duality can similarly be seen to arise from such a modular
map.
Further insight is obtained by using the set of boundary condition
basis vectors given in eq. (\ref{basis2}):
\begin{eqnarray}
v_1=1&=&\{\psi^\mu,\
\chi^{1,\dots,6},y^{1,\dots,6}, \omega^{1,\dots,6}| \nn\\
& & ~~~\bar{y}^{1,\dots,6},\bar{\omega}^{1,\dots,6},
\bar{\eta}^{1,2,3},
\bar{\psi}^{1,\dots,5},\bar{\phi}^{1,\dots,8}\},\nn\\
v_2=S&=&\{\psi^\mu,\chi^{1,\dots,6}\},\nn\\
v_{3}=z_1&=&\{\bar{\phi}^{1,\dots,4}\},\nn\\
v_{4}=z_2&=&\{\bar{\phi}^{5,\dots,8}\},\nn\\
v_{5}=z_3&=&\{\bar{\psi}^{1,\dots,4}\},\nn\\
v_{6}=z_0&=&\{\bar{\eta}^{0,1,2,3}\},\nn\\
v_{7}=b_1&=&\{\chi^{34},\chi^{56},y^{34},y^{56}|\bar{y}^{34},
\bar{y}^{56},\bar{\eta}^0,\bar{\eta}^{1}\},\label{basis2}
\end{eqnarray}
The models generated by the basis (\ref{basis2})
preserve $N=2$ space--time supersymmetry,
as the single supersymmetry breaking vector $b_1$
is included the basis. 

The novel feature in the basis of eq. (\ref{basis2}) is that the
spacetime vector bosons that are obtained in the untwisted Neveu--Schwarz
sector generate an $SO(12)\times SO(8)^4$ gauge symmetry at the $N=4$ level,
{\it i.e.} prior to the inclusion of the vector $b_1$. The lattice
$SO(12)$ symmetry
arises from the internal right--moving degrees of freedom,
$\{{\bar y},{\bar\omega}\}^{1,\cdots, 6}$, at the enhanced symmetry point,
whereas each of the
$SO(8)^4$ symmetries arise from the four sets of worldsheet fermions
$\{{\bar\psi}^{1,\cdots, 4}\};
\{{\bar\eta}^{0,1,2,3}\};
\{{\bar\phi}^{1, \cdots, 4} \};
\{{\bar\phi}^{1,\cdots, 6}\}$.
With the the basis given in eq. (\ref{basis2}), the
GUT $SO(10)$ in $N=1$ models, or $SO(12)$ in $N=2$ models,
is generated from vectors bosons in the purely anti--holomorphic
sectors
\beqn
G=\{& z_0,z_1,z_2,z_3, \nonumber\\
    & z_0+z_1,z_0+z_2,z_0+z_3,z_1+z_2,z_1+z_3,z_2+z_3~~\}. 
\label{gaugesectors}
\eeqn
The resulting gauge symmetries depend on the choices of GGSO projection
coefficients and have been classified in ref. \cite{fkrneq2}. They contain
the two cases $SO(12)\times E_8\times E_8$ and $SO(12)\times SO(16)\times
SO(16)$ that are distinguished by the choice of the phase
$$\cc{z_0}{z_1}=\pm1.$$
The $SO(12)\times SO(16)\times SO(16)$ gauge symmetry
is realised with the GGSO projection coefficients taken as
\beq
\cc{z_0}{z_1}=~~
\cc{z_0}{z_3}=~~
\cc{z_1}{z_2}=
\cc{z_0}{z_2}=
-\cc{z_1}{z_3}=
-\cc{z_2}{z_3}=-1,
\label{so16phases}
\eeq
where the enhancing vector bosons are obtained from the sectors $z_2$ and
$z_3$. 
As discussed above, the basis vector $b_1$ breaks $N=4$ spacetime supersymmetry
to $N=2$, and reduces gauge symmetry arising from the $0$--sector to
\beq
\left[SO(8)\times SO(4)\right]_{\cal L}
\times \left[SO(8)_3\times SO(4)\times SO(4)\right]_O\times
\left[SO(8)_1\times SO(8)_2\right]_H.
\eeq
For the choice given in eq. (\ref{so16phases}) the vector $b_1$ breaks
the $\left[SO(16)\right]_O\rightarrow \left[SO(12)\times SO(4)\right]_O
\equiv\left[SO(12)\times SU(2)_0\times SU(2)_1\right]_O$.
The realisation of the spinor--vector duality in this
model is now discussed.
First, consider the choice of additional phases given by: 
\beq
\cc{b_1}{1,z_0}=-\cc{b_1}{S,z_1,z_2,z_3}=-1~.
\label{vectorialphasechoice}
\eeq
In this case the model contains 2 multiplets in the
$(1,2_L+2_R,12,1,2,1)$ and 2 in the
$(8,2_L+2_R,1,2,1,1)$ representations of
$\left[SO(8)\times SO(4)\right]_{\cal L}
\times \left[ SO(12)\times SU(2)_0\times SU(2)_1\right]_{O}\times
\left[SO(16)\right]_H$.
the sectors giving rise to the vectorial 12 representation
of $SO(12)$ are the sectors $b_1$ and $b_1 + z_3$, where the sector $b_1$
produces the $(1,2,2)$ representation and the sectors $b_1+z_3$
produces the $(8_{_S},1,1)$ under the decomposition
$\left[SO(12)\right]_O
\rightarrow \left[SO(8)\times SO(4)\right]_O\equiv
\left[SO(8)\times SU(2)\times SU(2)\right]_O$.
All other states are projected out.
Therefore, there are a total of eight multiplets in the vectorial
representation of the observable $SO(12)$ in this model, which 
also transform as doublets of the observable $SU(2)_1$.

The second choice of GGSO phases is given by
\beq
\cc{b_1}{1,z_0,z_1}=-\cc{b_1}{S,z_2,z_3}=-1
\label{spinorialphasechoice}
\eeq
This choice produces a model with 2 multiplets in the $(1,2_L+2_R,32,1,1,1)$,
and 2 in the $(1,2_L+2_R,1,1,2,16)$, representations of
$\left[SO(8)\times SO(4)\right]_{\cal L}\times
\left[SO(12)\times SU(2)_0\times SU(2)_1\right]_O\times
\left[SO(16)\right]_H$.
In this case the the sectors giving rise to the spinorial 32 representation
of $\left[SO(12)\right]_O$ are the sectors $b_1+z_0$ and $b_1 + z_3+ z_0$,
where the sector $b_1+z_0$
produces the $(8_{_V},2,1)$ representation and the sectors $b_1+z_3+z_0$
produces the $(8_{_C},1,2)$ under the decomposition
$\left[SO(12)\right]_O\rightarrow
\left[SO(8)\times SO(4)\right]_O\equiv
\left[SO(8)\times SU(2)\times SU(2)\right]_O$.
The sectors producing the vectorial 16 representation
of the $SO(16)_H$ gauge group are the sectors $b_1$ and $b_1 + z_2$,
where the sector $b_1$
gives rise to the $(8_{_V},1)$ representation and the sector $b_1+z_2$
gives rise to the $(1,8_{_C})$ representation
under the decomposition
$\left[SO(16)\right]_H\rightarrow
\left[SO(8)_1\times SO(8)_2\right]_H$.
The hidden 16 representations
transform as doublets of the observable $SU(2)_1$ group.
All other states are projected out.
There are a total of eight multiplets in the spinorial 32
representation of the observable $\left[SO(12)\right]_O$ in this model.
The transformation between the two models,
(\ref{vectorialphasechoice}) and (\ref{spinorialphasechoice}),
is induced by the discrete GGSO phase change
\beq
\cc{b_1}{z_1}=+1~~\rightarrow~~\cc{b_1}{z_1}=-1
\label{dualitymap}
\eeq
What is crucial, however, is the role played by the basis vector
$z_0$ in eq. (\ref{basis2}). This basis vector induces a map between
the sectors that produce the spinorial states to those that produce
vectorial states. Its role in this respect is similar to that performed
by the $x$--vector in the basis of eq. (\ref{basis}), which induces the
co--called $x$--map of refs. \cite{fermionmasses, so10c10, cfkr}.
The $z_0$ basis vector is the analogue of the $S$ basis vector on the
supersymmetric side of the heterotic--string. In the case of $N=1$
models with enhanced $E_6$ symmetry, it acts as a spectral flow operator
that mixes the different components of the $E_6$ representations.
In the models in which $E_6$ is broken to $SO(10)\times U(1)$,
it induces the map between the spinor--vector dual models \cite{ffmt}.
The $z_0$ action is a second example of what is called here a ``modular
map''.

The two examples, the $S$--map and the $z_0$ map, are mere two examples
of a much richer symmetry structure. This much richer symmetry structure
is a topic of much interest in symmetries that underlie 24 dimensional
lattices. What role this symmetry structure plays in the phenomenological
properties of string theory is yet to be unravelled. The two examples
above clearly demonstrate its potential, the first being SUSY
phenomenology, whereas the second was investigated in the
context of low scale $Z^\prime$ phenomenology. An embryonic
attempt to link this rich symmetry structure to the phenomenological
free fermionic constructions was discussed in ref. \cite{panosleech}.
In the next section, I discuss how a similar ``modular map''
plays a role in the construction of non--supersymmetric
heterotic--string vacua that are compactifications of the
tachyonic string vacua in 10 dimensions, alluded to in figure
\ref{mtheorypicture}.

\section{Modular maps and non--supersymmetric string vacua}\label{nonsusy}

Since its advent in the mid nineteen eighties string phenomenology studies
have mainly focused on supersymmetric string vacua.
String theory, however, also gives rise to non--supersymmetric
ten dimensional vacua that may be tachyonic or non--tachyonic
\cite{tendnonssvacua}.
Supersymmetric string vacua are stable, whereas the non--supersymmetric
configurations are generically unstable. It is incumbent to understand their
role, in particular, in the early universe and the dynamics of string
vacuum selection. A good starting point for this exploration is the
$E_8\times E_8$ heterotic--string in ten dimensions.
The tachyon free $SO(16)\times SO(16)$
heterotic--string in ten dimensions is obtained as
an orbifold.
In the free fermion formulation, 
the $E_8\times E_8$ and $SO(16)\times SO(16)$ models are 
defined in terms of a common set of basis vectors 
\beqn
v_1={\bf1}&=&\{\psi^\mu,\
\chi^{1,\dots,6}| \overline{\eta}^{1,2,3},
\overline{\psi}^{1,\dots,5},\overline{\phi}^{1,\dots,8}\},\nonumber\\
v_{2}=z_1&=&\{\overline{\psi}^{1,\dots,5},
              \overline{\eta}^{1,2,3} \},\nonumber\\
v_{3}=z_2&=&\{\overline{\phi}^{1,\dots,8}\}.
\label{tendbasisvectors}
\eeqn
The spacetime supersymmetry generator 
is given by the combination 
\beq
S={\bf1}+z_1+z_2 = \{{\psi^\mu},\chi^{1,\dots,6}\}. 
\label{tendsvector}
\eeq
The GGSO phase $\cc{z_1}{z_2}=\pm1$ selects between 
the $E_8\times E_8$ or $SO(16)\times SO(16)$ heterotic--string vacua
in ten dimensions. The relation in eq. 
(\ref{tendsvector}) does not hold in lower dimensions, which entails that
in lower dimensions the projection of the supersymmetry generator is not
correlated with the breaking $E_8\times E_8\rightarrow SO(16)\times SO(16)$.
Compactifications of the
$SO(16)\times SO(16)$ heterotic--string model to 
four dimensions provide a basis for phenomenological studies
of non--supersymmetric heterotic--string vacua, which may, however,
contain tachyons \cite{s1616pheno}.
One can look for configurations of GGSO phases for
which all the tachyonic states are projected out. However, in this
respect, it possible similarly to start from a tachyonic ten dimensional
string vacuum and project all the tachyonic modes in the lower dimensional
theory. In terms of the modular maps that are our interest here, it is noted
that the $E_8\times E_8$ and $SO(16)\times SO(16)$ utilise the same modular
map. Namely, the $S$--vector.

Construction of vacua that descends from the tachyonic ten dimensional
vacua amounts to removing the $S$--vector from the basis generating the
models \cite{spwsp, fmp}. This can be achieved by removing the $S$--vector
entirely from the basis, or by augmenting it with four periodic worldsheet
fermions $\{{\bar\phi}^{1,...,4}\}$, defining
${\tilde S}=\{ \psi^{1,2}, \chi^{1,...,6}|{\bar\phi}^{1,...,4}\}$.
We then obtain a general $S\leftrightarrow {\tilde S}$ map, referred
to as the ${\tilde S}$--map, which is similar to the modular maps
discussed earlier, namely it is a map induced by a grouping of four
periodic worldsheet fermions. In the case of the $\tilde S$--map, the
map is induced between supersymmetric vacua and non--supersymmetric
vacua that are compactifications of the tachyonic ten dimensional vacua.
It is then of interest to explore the similarities and differences between
the different classes of models, both from the phenomenological
characteristics as well as structural. For example it is observed
that $\tilde S$--models can only be phenomenologically viable,
at least within this class, if the gauge symmetry is broken to
the Standard Model gauge symmetry. On the other hand, it is observed
that non--supersymmetric vacua, whether tachyonic or non--tachyonic,
exhibit an oscillatory behaviour of the massive spectrum between
bosonic and fermionic states, and that divergences in tachyonic
amplitudes arise solely due to the tachyonic state. That is, the
contribution to the amplitudes of the massless and massive modes
in the spectra of the models exhibit the same soft ultraviolet behaviour
of the tachyon free cases. Another question of interest is the
excess of massless fermionic versus massless bosonic states in the
models that determine the sign of the cosmological constant, where
models with $N_f^0-N_b^0>0$ may produce vacua with positive cosmological
constant. In this exploratory spirit we can search for string vacua with
extreme spectral characteristics, {\it e.g.} string vacua that have
no massless fermionic states, which are dubbed type 0 models, and those without
massless twisted bosonic states that are dubbed as type $\overline{0}$
vacua \cite{fmptype0}.
It is interesting that in both cases a form of misaligned
supersymmetry is present, partially explaining the mild behaviour
of string amplitudes even in the case of non--supersymmetric vacua.

\section{Conclusions}\label{conc}

The objective of mathematical modelling of experimental observations
is to minimise the number of arbitrary parameters required to
describe the experimental data. 
The experimentally observed data in the sub--atomic domain
strongly favours the realisation of grand unification structures
in nature, which reduces the number of ad hoc parameters in the
gauge and matter sectors of the Standard Model.
Whether the Standard Model is all there is, or whether physics
beyond the Standard Model is just around the corner, it is clear
that fundamental insight into the physical parameters can
only be gained by fusing it with gravity.
String theory is a perturbatively self--consistent theory of
quantum gravity and provides the arena to explore the gauge \& gravity
unification. String phenomenology aims to connect between string theory
and observational data. It is important to acknowledge that string
phenomenology is still at its initial stage of development and it may take
more than one lifetime, perhaps many lifetimes, to appreciate whether the
promise of string theory can be realised, or whether it is yet another vain
attempt at the construction of a tower of babel. The passing of this
judgement may occupy physicists throughout the third millennium. It will
not be the first occasion in history where decisive judgement had to await
nearly two millennia. Aristarchus of of Samos proposed a heliocentric model
of the solar system in the 3rd Century BC, but judgement of this
proposal had to await the development of observational instruments by Galileo
in the 17th Century that provided the decisive evidence.

String theory gives rise to a vast space of a priori possibly viable solutions,
which are being
studied using both worldsheet and effective field theory techniques.
However, the relation between the two approaches is fairly well understood
only in special cases with (2,2) worldsheet supersymmetry and the
so--called standard embedding. The more prevalent case with (2,0)
worldsheet supersymmetry is still mostly obscured. String theory, however,
exhibits duality relations between different string vacua that may
be used as a tool to probe the moduli spaces of string 
compactifications, in the effective field theory limit. Mirror symmetry
is the best known among those. It was initially observed in worldsheet
string compactifications and seen to have profound implications
on its effective field theory limits. Spinor--vector duality, discussed
in this talk, is akin to mirror symmetry, but whereas mirror symmetry arises
from exchanges of moduli of the internal three dimensional complex
manifold, spinor--vector duality arises from exchanges that correspond to
the gauge bundle moduli. As discussed herein, it is anticipated that the
spinor--vector duality is a mere example of a much wider symmetry
structure that is induced by ``modular maps'' and examples of what is
meant by such modular maps were discussed. Furthermore, the modular
maps have profound phenomenological implications that are relevant for
BSM phenomenology. Self--duality under the spinor--vector duality facilitates
the construction of string models that allow for an extra $U(1)$ gauge
symmetry to remain unbroken down to low scales. The particular
extra $U(1)$ symmetry in the string derived model implies the existence of
vector--like quarks and leptons at the $U(1)_{Z^\prime}$ breaking scale,
and may therefore impact precision measurements of the Standard Model
parameters, while evading direct searches.

\section*{Acknowledgements}
I would like to thank
Martin Hurtado Heredia,
Stefan Groot--Nibbelink,
Viktor Matyas,
Benjamin Percival and
John Rizos for collaboration, and the organisers for the invitation to
present this work at the BSM--2021 conference. 

\bibliographystyle{unsrt}

\end{document}